\documentclass[aps,prb,twocolumn,groupedaddress,showpacs]{revtex4}
\usepackage{dcolumn}
\usepackage{epsfig}
\usepackage{graphicx}
\usepackage{color}

\begin{document}

\title{Coexistence pressure for a martensitic transformation from theory and experiment: 
 revisiting  the bcc-hcp transition of iron under pressure}
%{Iron under pressure: bcc-hcp equilibrium coexistence revisited}

\author{N.~A. Zarkevich,$^{1}$  and D.~D. Johnson$^{1,2}$}
\email{zarkev@ameslab.gov;   ddj@ameslab.gov}
\affiliation{$^{1}$The Ames Laboratory, U.S. Department of Energy, Ames, Iowa 50011-3020 USA;}
\affiliation{$^{2}$Materials Science \& Engineering, Iowa State University, Ames, Iowa 50011-2300 USA.}

\begin{abstract}
%In theory, 
The coexistence pressure of two phases  is a well-defined point at fixed temperature.
In experiment, however, due to non-hydrostatic stresses and a stress-dependent potential energy barrier, different measurements yield different ranges of pressure with a hysteresis.
Accounting for these effects,  we propose an  inequality  for comparison of the theoretical value 
to a plurality of measured intervals. 
We revisit decades of pressure experiments on the $bcc \leftrightarrow hcp$ transformations in iron, which are sensitive to non-hydrostatic conditions and sample size. 
From electronic-structure calculations, we find a $bcc \leftrightarrow hcp$ coexistence pressure of $8.4~$GPa.  We construct the equation of state for competing phases under hydrostatic pressure, compare to experiments and other calculations, and address the observed pressure hysteresis and range of onset pressures of the nucleating phase.
\end{abstract}


%%% PACS numbers
%02.70.-c	Computational techniques; simulations
%05.10.-a	Computational methods in statistical physics and nonlinear dynamics
%05.50.+q	Lattice theory and statistics (Ising, Potts, etc.) (see also 64.60.Cn Order-disorder transformations, and 75.10.Hk Classical spin models)
%05.70.-a	Thermodynamics
%05.70.Fh	Phase transitions: general studies
%61.50.-f	Structure of bulk crystals
%61.66.Bi	Elemental solids
%64.	Equations of state, phase equilibria, and phase transitions (see also 82.60.-s Chemical thermodynamics)
%64.60.Bd	General theory of phase transitions
%64.60.Ej	Studies/theory of phase transitions of specific substances (for phase transitions in ferroelectric and antiferroelectric materials, see 77.80.B-)
%64.70.-p	Specific phase transitions
%64.70.K-	Solid-solid transitions (see also 61.50.Ks Crystallographic aspects of phase transformations; pressure effects; 75.30.Kz and 77.80.B- for magnetic and ferroelectric transitions, respectively; for materials science aspects, see 81.30.-t)
%75.30.-m	Intrinsic properties of magnetically ordered materials
%75.47.Np	Metals and alloys
%81.05.Bx	Metals, semimetals, and alloys

\maketitle
%\linenumbers

\section{Introduction}
Knowledge of the equation of state is crucially important in materials science and engineering, metallurgy, geophysics, and planetary sciences. However, equilibrium coexistence of phases during a pressure-induced martensitic transformation is extremely difficult to realize experimentally, and most shock and anvil cell experiments contain various amounts of a non-hydrostatic, anisotropic stress. 
Hence, an improved understanding of transformations arises when we can better compare idealized theoretical results to realistic experimental data. A long-studied case is iron (Fe), our focus below, but the results remain quite general.

{\par }Iron is the most stable element produced by nuclear reactions at ambient pressure, and one of the most abundant elements in the Earth. Thus, magneto-structural transformations 
\cite{1,2,3,4,200,5,6,7,8,12,11,14,23,27,22,9,10,13,15,16,17,18,19,20,160,24,25,26} 
%\cite{1}--\cite{26}
and high-pressure states %\cite{28}--\cite{67} 
\cite{28,29,30,31,32,33,34,35,36,37,56,38,B8,39,40,41,42,43,44,45,46,47,48,49,50,51,52,53,54,55,57,58,59,60,61,62,63,64,65,66,67}
in iron attract enormous interest, especially in geophysics because iron is a primary constituent of the Earth's core, \cite{28}$^-$\cite{181} 
many meteorites, %\cite{181}--\cite{199} 
\cite{181,AgeIronMeteorites1930,185,189,190,193,194,196,197,199}
and, due to its properties and availability, most steels. \cite{steel} 
At low pressure $P$ and temperature $T$, the $\alpha$-phase of iron is a ferromagnet (FM) with the body-centered cubic (bcc) structure. At higher pressures, iron transforms to the $\varepsilon$-phase with hexagonal close-packed (hcp) structure of higher density that is non-magnetic or weakly anti-ferromagnetic. This transformation is martensitic, \cite{3} and the bcc-hcp equilibrium coexistence pressure is difficult to determine unambiguously  experimentally (Table 1). 

\begin{figure*}[ht]
\begin{center}
\includegraphics[scale=1]{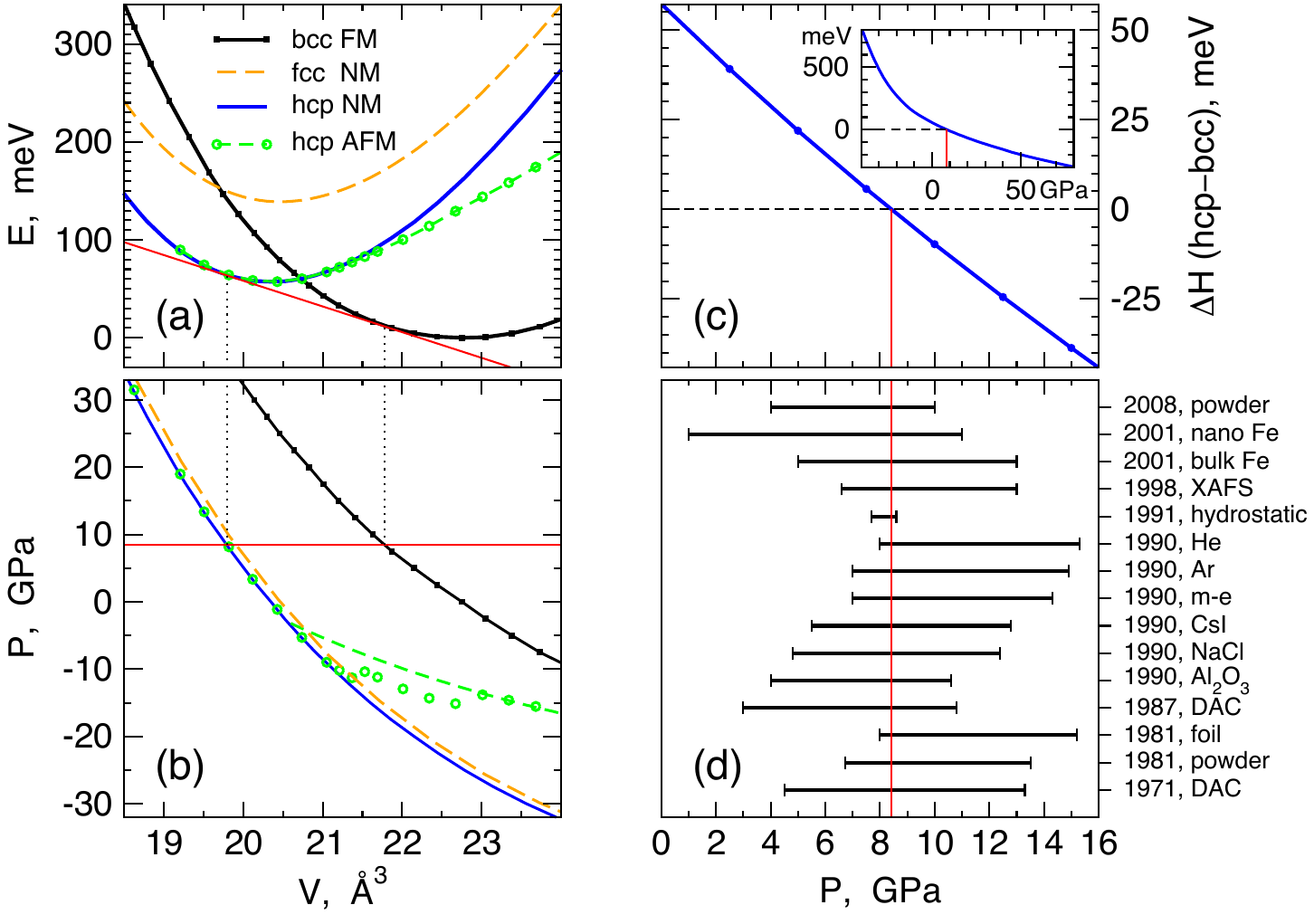}
\caption{\label{fig1}
Figure 1. (a) Energy [meV/atom] relative to bcc at $0~$GPa and (b) pressure [GPa] versus volume [\AA$^3$/cell] for hydrostatically relaxed 2-atom unit cells of bcc FM (black), hcp NM (blue) and AFM (green), and fcc NM iron (orange); with DFT values (dots) and least-squares fit to the Birch-Murnaghan EoS (lines).
Common tangent construction (red line) yields $P_0=8.42 \,$GPa.  Vertical dotted lines are guides to the eye. (c) Enthalpy difference [meV/Fe] between hcp and bcc phases versus pressure [GPa] (inset shows a larger range). (d) Comparison of the calculated $P_0=8.42 \,$GPa (vertical red line) with experimental data from Table 1, represented by 
[$P_{end}^{\varepsilon \rightarrow \alpha} - P_{start}^{\alpha \rightarrow \varepsilon}$] horizontal segments. 
Except for the 1991 hydrostatic experiment \cite{8}, most diamond anvil cells \cite{3,5} provided uniaxial or highly anisotropic pressure.}
\end{center}
\end{figure*}

{\par}A martensitic transformation between bcc ($\alpha$) and hcp ($\varepsilon$) phases can be characterized by four pressures (Table 1 and Fig.~\ref{fig1}): a \emph{start} and \emph{end} pressure of direct $\alpha \rightarrow \varepsilon$  ($P_{start}^{\alpha \rightarrow \varepsilon}$, $P_{end}^{\alpha \rightarrow \varepsilon}$) and reverse  $\varepsilon \rightarrow \alpha$ ($P_{start}^{\varepsilon \rightarrow \alpha}$, $P_{end}^{\varepsilon \rightarrow \alpha}$) transformations. 
Because martensitic stress is present in the anisotropic hcp phase but not in the isotropic bcc phase,  
we suggest the inequality
\begin{equation} %(1)
\label{eq1}
P_{end}^{\varepsilon \rightarrow \alpha} < P_0 < P_{start}^{\alpha \rightarrow \varepsilon} 
      \end{equation}
for the $\alpha - \varepsilon$ equilibrium coexistence pressure $P_0$  and the observed hysteresis, rather than an inaccurate simple average \cite{3}
\begin{equation} %(2)
\label{eq2}
 P_{start}^{avg.} = \frac{1}{2} \left( P_{start}^{\alpha \rightarrow \varepsilon} + P_{start}^{\varepsilon \rightarrow \alpha} \right) .	
\end{equation}
While shock and anvil-cell (AC) pressure experiments give different averages (\ref{eq2}), they satisfy the more appropriate inequality (\ref{eq1}), see Table~1 and Fig.~1d.
Additionally, we calculate the hydrostatic equation of state (EoS) of $\alpha$ and $\varepsilon$ Fe, determine $P_0$ via common-tangent construction, which should be thermodynamically relevant to purely hydrostatic (equilibrium) AC experiments, and compare the result to experiment.

\begin{table*}
\begin{center}
\begin{tabular}{rcccccccc}
\hline
Ref. &	Year	& $P_{start}^{\alpha \rightarrow \varepsilon}$	
       & $P_{end}^{\alpha \rightarrow \varepsilon}$ 
      &	$\Delta P^{\alpha \rightarrow \varepsilon}$	
     & $P_{start}^{\varepsilon \rightarrow \alpha}$	
    & $P_{end}^{\varepsilon \rightarrow \alpha}$	
    & $\Delta P^{\varepsilon \rightarrow \alpha}$	
    & Expt.\\
\hline
\cite{1} & 1956	& 13.1	& & & & & & shock\\
\cite{2} & 1961	& 13.3	& & & & & & resistance\\
\cite{3} & 1971	& 13.3	& 16.3	& 3	& 8.1	& 4.5	& 3.6	& AC\\
\cite{4} & 1981	& 13.52	& 15.27	& $<$2	& 9.2	& 6.74	& 2.5	& powder\\
\cite{4} & 	& 15.21	& 15.47	& $<$1	& 10.23	& 8.5(6) & 2	& foil\\
\cite{5} & 1987	& 10.8	& 21	& $\approx $10	& 15.8	& 3	& 13	& Au\\
\cite{6} & 1990	& 10.6	& 25.4	& 14.8	& 16	& 4	& 12	& Al$_2$O$_3$\\
\cite{6} & 	& 10.7	& 21.6	& 10.9	& 16.2	& 3.7	& 12.5	& Au\\
\cite{6} & 	& 12.4	& 17.8	& 5.4	& 12.2	& 4.8	& 7.4	& NaCl\\
\cite{6} & 	& 12.8	& 17.2	& 4.4	& 11.8	& 5.5	& 6.3	& CsI\\
\cite{6} & 	& 14.3	& 17.5	& 3.2	& 11.9	& 7	& 5	& m-e\\
\cite{6} & 	& 14.9	& $<$15.9	& 0.5	& $<$11	& $>$7	& $<$4	& Ar\\
\cite{6} & 	& 15.3	& 15.3	& 0.1	& 10.6	& 8.0(6)	& 2	& He\\
\cite{8} & 1991	& 8.6	& 23	& $\approx $14	 & [9.5]	& 7.7	& 3.6	& hydrostatic\\
\cite{11} & 1998 & 13.0 & 18.6 & $\approx $5.6	 & [10.3]	& 6.6	& 7.4	& XAFS\\  %\cite{11,12}
\cite{14} & 2001 & 13	& 17	& $\approx $4	& 8	& 5	& 3	& bulk\\
\cite{14} & 	& 11	& 14	& $\approx $3	& 7	& 1	& 6	& nano-Fe\\
\cite{23} & 2005 & 14	& 16	& 2.4		& 	& 	& & AC\\
\cite{27} & 2008 & 10	& 22	& $\approx $12	& 8	& 4	& 4	& powder\\
\hline
\end{tabular}
\caption{\label{table1}
Start and end pressures [GPa] with width $\Delta P = |P_{end} - P_{start} |$ 
for iron bcc ($\alpha $) -- hcp ($\varepsilon$) direct and inverse transformations. 
Type of experiment (Expt.) specifies shock or anvil cell (AC), form of sample (bulk, foil, powder), 
or pressure medium (He, Ar, “m-e” for methanol-ethanol, etc.) in the AC. 
The $P_{1/2}^{\varepsilon \rightarrow \alpha}$ values at half-transition (50\% bcc + 50\% hcp) are in the square brackets 
[$P_{start}^{\varepsilon \rightarrow \alpha}$ column].  }
\end{center}
\end{table*}

\section{Background}
\emph{Previous Experiments: }Shock and AC pressure experiments are the major approaches to measure pressure-induced transformations, although hydrostatic conditions are often difficult to assess. 
Experimental onset (start) and final (end) pressures for $\alpha \rightarrow \varepsilon$ and  $\varepsilon \rightarrow \alpha$ transformations are summarized in Table~1, which show a large span and the reason to revisit this issue. 

{\par } For completeness, we highlight the experiments and their outcome for iron. 
Bancroft \emph{et al.} \cite{1} studied propagation of compressive waves generated by high explosive in Armco iron and reported a polymorphic transition at 13.1 GPa. Balchan and Drickamer \cite{2} used a high-pressure electrical resistance cell and found a sharp rise in resistance of iron at 13.3 GPa. Giles \emph{et al.} \cite{3} showed that this bcc-hcp transformation is martensitic; their estimate of $P_0$ by 
$P_{start}^{avg.}=10.7 \pm 0.8 \,$GPa differs from the earlier reported 
$P_{start}^{\alpha \rightarrow \varepsilon}=13 \,$GPa, often quoted as the martensitic start pressure. 
Mao, Bassett, and Takahashi \cite{200} performed XRD measurements of lattice parameters of iron at $23^{\circ}$C at pressures up to 30 GPa, and suggested a bcc-hcp shear-shuffle model. (Their Fig.~3 is reproduced in Ref.~\cite{5}.) 
Bassett and Huang \cite{5} applied a non-hydrostatic pressure with an uncontrolled shear strain (known to produce pressure self-multiplication) \cite{201} and confirmed an atomic mechanism \cite{200} of the bcc-hcp transition, but omitted discussion of changes in volume and magnetization in their shear-shuffle model.
Zou \emph{et al.} \cite{4} used solid He as the pressure medium in their diamond AC (DAC) experiments on iron (99.95 wt.\% Fe) powder pressed into a plate and on a folded section of a 10 micron foil; they pointed at the uniform non-hydrostatic stress as a possible cause of differing data.

{\par} Importantly, transition pressure estimates depend on how hydrostatic the applied stress is and sample size.  For example, Bargen and Boehler \cite{7} found that the pressure interval of the forward bcc$ \rightarrow $hcp transition increases with increasing non-hydrostaticity (transition pressures and hysteresis width change systematically with the shear strength of the pressure medium). \cite{6} 
The best pressure medium is a superfluid; a good one is a gas or a fluid with a low viscosity; 
 the worst one is a viscous fluid or a solid. 
Due to grain boundaries \cite{supersolidHe} 
and %produced by vibrations 
melting-freezing waves, \cite{He4waves}
solid helium (He) can behave as a superfluid. \cite{superHe}  %the best pressurizing medium with zero viscosity.

{\par}
Taylor \emph{et al.} \cite{8} focused on the large hysteresis and used a DAC up to 24 GPa; as pressure is increased, Fe is fully converted to hcp at $P_{end}^{\alpha \rightarrow \varepsilon}=23 \,$GPa. Upon reducing  pressure, half of the hcp transforms to bcc by $P_{1/2}^{\varepsilon \rightarrow \alpha}=9.5 \,$GPa, while a small $\varepsilon$-Fe remnant  is present at $P_{end}^{\varepsilon \rightarrow \alpha}=7.7 \,$GPa. They report $P_{start}^{\alpha \rightarrow \varepsilon}$ values from 8.6 to 15 GPa \cite{8}.  
Using a radial diffraction DAC with infrared laser heating on Alfa Aesar (99.9\% pure) Fe powder ($10^{-5}\,$m particle size), Miyagi \emph{et al.} \cite{27} reported appearance of hcp at 10 GPa that fully converts near $22 \,$GPa, while bcc appears at 8 GPa during decompression.
Jiang \emph{et al.} \cite{14} studied grain-size and alloying effects on the transition pressure, finding that $P_{start}^{\alpha \rightarrow \varepsilon}$ shifts from 13 GPa in bulk to 11 GPa in nano-crystalline samples (15 nm average grain size with a range of 10-30 nm). 
Wang, Ingalls, and Crozier \cite{12} performed an XAFS study at 23$^\circ$C  up to 21.5 GPa; a mixed-phase region was found between $P_{start}^{\alpha \rightarrow \varepsilon}=13$ and $P_{end}^{\alpha \rightarrow \varepsilon}=20 \,$GPa, and between $P_{start}^{\varepsilon \rightarrow \alpha}=15$ and $P_{end}^{\varepsilon \rightarrow \alpha}=11$ GPa. 
Later, Wang and Ingalls \cite{11} used XAFS with a sintered boron-carbide anvil cell to measure lattice constants and bcc abundance versus P, and reported  $P_{start}^{\alpha \rightarrow \varepsilon}=13 \,$GPa and $6.6  \le P_{end}^{\varepsilon \rightarrow \alpha}   \le 8.9 $ GPa.
Using \emph{in situ} EXAFS measurements and nanosecond laser shocks, Yaakobi \emph{et al.} \cite{22} detected hcp phase and claimed that the $\alpha \rightarrow \varepsilon$ transition can happen very quickly. 

{\par} Finally, the change of magnetization along a transition path is important, where there is an abrupt 8--10\% volume decrease  at the transition state \cite{5,23}.  Baudelet \emph{et al.} \cite{23} combined x-ray absorption spectroscopy (XAS) and  x-ray magnetic circular dichroism (XMCD) on a sample in a CuBe DAC and found a transition at 14 GPa, with a $2.4 \pm 0.2 \,$GPa width of the local structural transition and a $2.2 \pm 0.2 \,$GPa width of the magnetic one; they suggest that the magnetic moment collapse lies at the origin of the structural transition, and slightly precedes the structural one. 

{\par} \emph{Previous Theory Results: } Former bcc-hcp equilibrium pressure calculations provided values of 13.1 GPa \cite{15}, 10.5 GPa \cite{20} and 10 GPa \cite{160}, in apparent agreement with the experimental values of  $P_{start}^{\alpha \rightarrow \varepsilon}=13 \,$GPa \cite{1,2,3} and 
$P_{start}^{avg.}=10.7 \pm 0.8 \,$GPa \cite{3}.  However, those calculated pressures disagree with later experimental data (Table 1). 
	Using \emph{ab initio} molecular dynamics (MD), Belonoshko \emph{et al.}  \cite{117,202,203,204,205,206} considered shear at the Earth's core conditions \cite{93,94} and constructed an EoS for $\alpha$ \cite{38, 205} and $\varepsilon$  \cite{206} Fe.  
	Wang \emph{et al.} \cite{19} studied nucleation of the higher-pressure hcp and fcc phases by classical MD simulations employing an embedded atom method (EAM) potential, and found that the transformation happens on a picosecond timescale; their calculated transition pressure is around 31-33 GPa for uniform \cite{19} and 14 GPa for uniaxial \cite{18} compression (but there is no magnetization in the EAM potential). 
	Caspersen \emph{et al.} \cite{24} showed that presence of a modest shear accounts for the scatter in measured transformation pressures, affecting the hysteresis. Johnson and Carter \cite{15} used a drag method in a rapid-nuclear-motion (RNM) approximation and obtained an unphysical discontinuous jump in atomic shuffle degrees of freedom, giving a very low bcc-hcp barrier; they found that bcc and hcp phases have equal enthalpies at the calculated pressure of 13.1 GPa. 

{\par}Liu and Johnson \cite{16} directly constructed the potential energy surface in a 2-atom cell for the shear-shuffle model \cite{5}, allowing changes of lattice constants and (continuous) atomic degrees of freedom; although hydrostatic pressure cannot produce shear, pressure does affect the potential energy surface and  barriers. They reported $\approx$$9$ GPa for bcc-hcp coexistence; the calculated kinetic barriers along the transition path were 132 meV/atom at 0 GPa with an estimated minimum (maximum) onset pressure of 9 (12.6) GPa, 119 meV/atom at 10.5 GPa  with a min (max) onset at 8.1 (13.8) GPa, and 96 meV/atom at 22 GPa with a min (max) onset of 6.6 (10.2) GPa. That is, there is an expected $3.6$ to $5.7$ GPa hysteresis width depending on kinetic pathway (and volume fluctuations). In addition, they showed that drag methods decouple degrees of freedom incorrectly, as  confirmed later by a proper  solid-solid nudged-elastic band method \cite{GSSNEB}.

{\par}Recently, Dupe \emph{et al.} \cite{207} reconsidered the transition mechanism within the same shear-shuffle model \cite{5}, but incorrectly fixed the volume at 71.5 bohr$^3$/atom (no moment collapse allowed), and used the RNM drag method to compare energies of three shuffling mechanisms at constant shear and volume. 
Friak an Sob \cite{20} in a 4-atom cell  considered non-magnetic (NM) and  antiferromagnetic (AFM) orderings along a pre\-defined path (which were almost degenerate); their energy-volume common-tangent gave coexistence $P_0$ at 10.5 GPa \cite{20}.

%Equilibrium coexistence pressure:

%\emph{Present Results: } 
\section{Present Results} 
To determine $P_0$ of equilibrium coexistence of FM bcc and NM hcp phases, we calculate volume $V$, energy $E$, and enthalpy $H=E+PV$ (Fig.~1) at various hydrostatic external pressures $P$. Each unit cell is fully relaxed at a given P. All atomic forces and all non-diagonal pressure components remain zero due to symmetry.  Diagonal pressure components are the same by symmetry in bcc and fcc phases, while their difference does not exceed 0.03 GPa in hcp. Magnetization of the FM bcc reduces with pressure and collapses to zero at $\approx$900 GPa; hcp magnetization is set to zero at all pressures. 

{\par }The slope of the common tangent to the $E(V)$ curves in Fig.~1a gives $P_0$ of 8.4 GPa (a more accurate result than in \cite{16}, where the focus was on transition barriers); this pressure gives zero enthalpy difference in Fig.~1c, and  is compared to all experiments in Fig.~1d.  The previously calculated values of 13.1 GPa \cite{15} and 10.5 GPa \cite{20} do not agree with all the experimental data, summarized in Table~1 and Fig.~1d. 

%Computational details:
{\par} To obtain these results, we used the Vienna ab initio simulation package (VASP) \cite{208,209,210} with generalized gradient approximation (GGA) \cite{211,212} and projector augmented-wave (PAW) potentials \cite{213, 214}.  We use 334.88 eV energy cutoff for the plane-wave basis with augmentation charge cutoff of 511.4 eV. The modified Broyden method \cite{215} is used for self-consistency. We carefully check convergence with respect to the number of $k$-points (up to $32^3$=32768) in the $\Gamma$-centered Monkhorst-Pack \cite{216} mesh within the tetrahedron method with Bl\"ochl corrections. Gaussian smearing with $\sigma=0.05 \,$eV with $16^3$=4096 k-points in the 2-atom cell is used for relaxation. 
%Exchange correlation:
 The role of the exchange correlation functional was considered in \cite{160,PRB79p085104}.  We use PBE-PAW-GGA to provide reasonable agreement with experiment for the lattice constants, compressibilities, and energies.  The expected systematic errors in the equilibrium lattice constants $\varepsilon (a) \le 1\%$, 
volume $\varepsilon (V) = [\varepsilon (a)]^3 \le 3\%$, and relative energies $\delta E \le 1\,$meV/atom give an estimate of the error in $P_0$ not exceeding 0.5 GPa. 

%Equation of state:
{\par}There are many EoS for solids \cite{217}. We fit our $E(V)$ data in Fig.~1a to the Birch-Murnaghan \cite{218, 219}  
\begin{equation} %(3)
\label{eqEV}
E(V)=E_0+\frac{9}{16} V_0 B_0 \left[f^3 B_0^\prime +2(1-2f)f^2 \right],			
\end{equation}
with $f=[(V/V_0 )^{2/3}-1]$.  For iron, the parameters are given in Table 2 for FM bcc at low pressure, and NM hcp at high pressure. Although hcp at lower pressure and density ($V>23 \,$\AA$^3$/cell) changes from NM to AFM,  their $E(V)$ curves at $V<21\,${\AA}$^3$ are almost degenerate. These values have some dependence on the range of fitted data, and are affected by the EoS functional form. 
As expected, calculated volume $V_0$ is reduced by 3\%  compared to experiment due to the standard DFT systematic error (i.e., 1\% in lattice constants).  This DFT error introduces a systematic 3\% error (0.25 GPa) in our bcc-hcp coexistence pressure. 

{\par} Our result for bcc iron is in agreement with the previous DFT calculations \cite{160,220}, with $B_0$ ranging from 171 to 194 GPa from EMTO, VASP, and Wien2K codes, which compare well with the assesses values of 195--205 GPa, see Table~3.1 on p.~47 in \cite{FizValues}.
	Our EoS coefficients for the hcp single crystal agree with previously calculated ones \cite{221, 222} at T=$0\,$K, summarized in Table~1 in \cite{221}. However, the experimentally assessed EoS for hcp martensite with $B_0$ of 166-195 GPa and B$_0^\prime$ of 4.3-5.3 differs from that calculated for a hcp single crystal (Table~2). 
This difference is expected because a martensite is a composite with both compressed and dilated regions. % with varying anisotropic stresses. 
Any non-homogeneous distortion increases energy, shifting up and distorting the $E(V)$ curve in Fig.~\ref{fig1}a.

\begin{table}
\begin{center}
\begin{tabular}{l|cl|c|c}
\hline
  & \multicolumn{2}{c}{$V_0$ } & $B_0$ & $B_0^\prime$\\
  & $\frac{\mbox{\AA}^3}{\mbox{cell}}$ & $\frac{\mbox{cm}^3}{\mbox{mol}}$ & GPa &   \\
\vspace{-4.2mm}
\\
\hline
bcc  FM	& 22.72	& 6.84	& 185	& 4.7\\
hcp  NM	& 20.34	& 6.13	& 293	& 4.5\\
hcp  AFM	& 19.94	& 6.004	& 140	& 3.9\\
% & & $\frac{\mbox{\AA}^3}{\mbox{cell}}$ & $\frac{\mbox{cm}^3}{\mbox{mol}}$ & GPa &   \\
\hline
\end{tabular}
\caption{\label{table2}
 Birch-Murnaghan EoS parameters for iron. }
\end{center}
\end{table}

%Discussion and Comparison to experiment:
%\emph{Discussion: } 
\section{Discussion}
Transformation from $\alpha$ (bcc) to $\varepsilon$ (hcp) iron is martensitic \cite{3}, and the hysteresis loop can be characterized by four pressures: $P_{start}^{\alpha \rightarrow \varepsilon}$, $P_{end}^{\alpha \rightarrow \varepsilon}$, $P_{start}^{\varepsilon \rightarrow \alpha}$, and $P_{end}^{\alpha \rightarrow \varepsilon}$. 
In experiment \cite{5,6}, $\varepsilon$-phase appears at $P_{start}^{\alpha \rightarrow \varepsilon}$ between 8.6 and 15.3 GPa, while $\alpha$-phase is fully converted above $P_{end}^{\alpha \rightarrow \varepsilon}$ between 14 and 25 GPa upon loading.  Whereas, upon unloading, $\alpha$-phase appears at $P_{start}^{\varepsilon \rightarrow \alpha}$ between 16 and 7 GPa and $\varepsilon$-phase  disappears below $P_{end}^{\varepsilon \rightarrow \alpha}$ between 8 and 1 GPa. 
Importantly, there is no strict inequality between $P_{start}^{\alpha \rightarrow \varepsilon}$ and $P_{start}^{\varepsilon \rightarrow \alpha}$ due to the martensitic stress distribution in the $\varepsilon$-phase. 

{\par}Our calculated $P_0$ of $8.4 \,$GPa is below $P_{start}^{\alpha \rightarrow \varepsilon}$ and above $P_{end}^{\varepsilon \rightarrow \alpha}$, see inequality (\ref{eq1}).  It agrees well with the experimental distribution of $P_{start}^{\alpha \rightarrow \varepsilon} \ge 8.6 \,$GPa and $P_{end}^{\varepsilon \rightarrow \alpha} \le 8.5 \,$GPa. 
	The observed $P_{start}^{\varepsilon \rightarrow \alpha}$ and $P_{end}^{\alpha \rightarrow \varepsilon}$ are highly affected by the martensitic stress within the hcp $\varepsilon$-phase. A martensitic transformation occurs between an isotropic (bcc) austenite and an anisotropic (hcp) martensite, which experience martensitic stress resulting in anisotropic distortions. In other words, there is little internal stress in austenite and large anisotropic internal stresses in martensite. However, martensitic stress is not taken into account in our calculation of the bcc-hcp equilibrium coexistence pressure $P_0$. Because hcp does not exist below $P_{start}^{\alpha \rightarrow \varepsilon}$ and $P_{end}^{\varepsilon \rightarrow \alpha}$, these values should not be affected by the martensitic stress in hcp  (though the transformation can be delayed due to an energy barrier), and can be used in the proper comparison to experiment. Hence, $P_0$ must be between $P_{start}^{\alpha \rightarrow \varepsilon}$ and $P_{end}^{\varepsilon \rightarrow \alpha}$, see inequality (\ref{eq1}).   These experimental ranges are compared with our calculated value of $P_0$ in Fig.~1d, with an excellent agreement between theory and experiment. 
	
{\par} Elsewhere we will present the energy barriers and transition states via a generalized solid-solid nudged-elastic band that incorporates both volume and magnetization collapse, needed for understanding of the observed abrupt magneto-volume effects. Change of magnetization of the transition state from FM to NM results in the pressure change by $\Delta P=24 \,$GPa. This calculated $\Delta P$ at the transition state agrees with the observed  bcc-hcp coexistence interval $[P_{end}^{\varepsilon \rightarrow \alpha}, P_{end}^{\alpha \rightarrow \varepsilon}]$. 
% MAYBE -BUT THIS IS A RANDOM COMMENT  %Chemical stabilization of the hcp phase at ambient conditions can result in a new family of steels with properties similar to those of the hcp cobalt or titanium (zirconium, hafnium) alloys. 

%Summary:
\section{Summary}
{\par} We provide a methodology for comparing idealistic theoretical predictions at hydrostatic pressure 
to realistic experiments with anisotropic stress, based on inequality \ref{eq1}. 
For the iron bcc-hcp equilibrium coexistence, our calculated pressure of 8.42 GPa is in agreement with available experimental data. Anisotropic internal stress in the hcp martensite, difference in volume between FM bcc and NM (and competing AFM) hcp iron near the transition state contribute to the spread of the experimentally assessed (non-equilibrium) bcc-hcp coexistence pressures, as well as to the uncertainty in the equation of state of hcp martensite. We emphasized the difference between a single crystal and a martensite and improved understanding of the available data for iron under pressure. Importantly, we suggested a universal  inequality (1), graphically illustrated in Fig.~1d, for proper comparison of the assessed and calculated pressures characterizing magneto-structural (martensitic) transformations in many materials.  

{\par}
%Acknowledgements:
\section*{Acknowledgements}
	We thank Graeme Henkelman and Iver Anderson for discussion. This work was supported  by the U.S. Department of Energy (DOE), Office of Science, Basic Energy Sciences (BES), Materials Science and Engineering Division.  This research was performed at the Ames Laboratory, which is operated for the U.S. DOE by Iowa State University under contract DE-AC02-07CH11358.

\section*{References}
%http://www.efm.leeds.ac.uk/~mark/ISIabbr/J_abrvjt.html
%\bibliography{Fe}

\end{document}